\documentclass{article}
\usepackage{ijcai25}

\usepackage{times}
\usepackage{soul}
\usepackage{url}
\usepackage[hidelinks]{hyperref}
\usepackage[utf8]{inputenc}
\usepackage[small]{caption}
\usepackage{subcaption}
\usepackage{graphicx}
\usepackage{amsmath,amssymb,amsfonts}
\usepackage{amsthm}
\usepackage{xcolor}
\usepackage{multirow}
\usepackage{array}
\usepackage{pifont}
\usepackage{booktabs}
\usepackage{algorithm}
\usepackage{algorithmic}
\usepackage[switch]{lineno}
\usepackage{lipsum}
\newcommand{\M}{CRWKV}

\title{Multi-View Learning with Context-Guided Receptance for Image Denoising}

\author{
\large{
Binghong Chen$^1$ \and
Tingting Chai$^{2}$\thanks{Corresponding author.} \and
Wei Jiang$^1$\and
Yuanrong Xu$^2$\and
Guanglu Zhou$^2$\And
Xiangqian Wu$^2$}\\
\vspace{1ex}
\affiliations
\large{$^1$School of Mathematics, Harbin Institute of Technology, China}\\
\large{$^2$Faculty of Computing, Harbin Institute of Technology, China}\\
\emails \large{math.cbh@stu.hit.edu.cn}, \large{ttchai@hit.edu.cn}
}

\usepackage[
    final,
]{changes}

\begin{document}

\maketitle

\begin{abstract}
Image denoising is essential in low-level vision applications such as photography and automated driving.
Existing methods struggle with distinguishing complex noise patterns in real-world scenes and consume significant computational resources due to reliance on Transformer-based models.
In this work, the Context-guided Receptance Weighted Key-Value (\M) model is proposed, combining enhanced multi-view feature integration with efficient sequence modeling.  The Context-guided Token Shift (CTS) mechanism is introduced to effectively capture local spatial dependencies and enhance the model's ability to model real-world noise distributions. Also, the Frequency Mix (FMix) module extracting frequency-domain features is designed to isolate noise in high-frequency spectra, and is integrated with spatial representations through a multi-view learning process. To improve computational efficiency, the Bidirectional WKV (BiWKV) mechanism is adopted, enabling full pixel-sequence interaction with linear complexity while overcoming the causal selection constraints.
The model is validated on multiple real-world image denoising datasets, outperforming the state-of-the-art methods quantitatively and reducing inference time up to 40\%. Qualitative results further demonstrate the ability of our model to restore fine details in various scenes. The code is publicly available at \url{https://github.com/Seeker98/CRWKV}.
\end{abstract}
 
\section{Introduction}
\begin{figure}[!t]
\centering
\includegraphics[width=\linewidth]{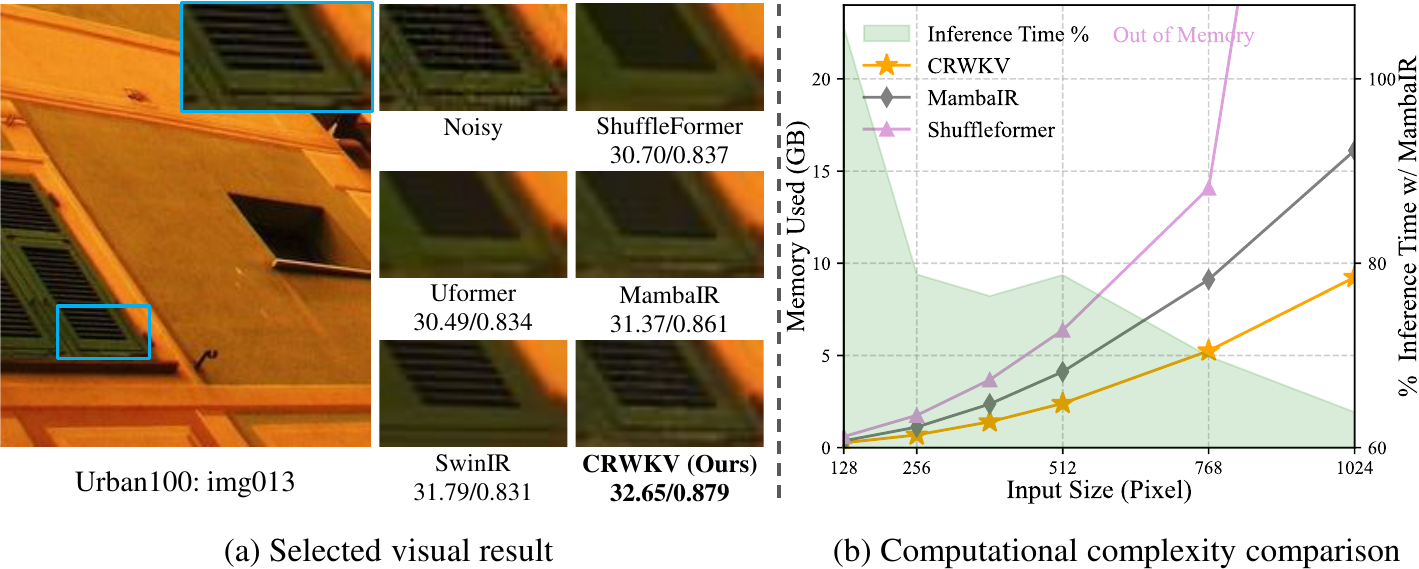}
\caption{Comparison of existing image denoising methods. (a) Visual results on Urban100 dataset: our method preserves fine details under complex noise pattern, while others suffer from information losses or artifacts. (b) Computational complexity across input scales for Transformer- and state space-based models.}

\label{fig:fig1}
\end{figure}

Images captured in real-world scenes are commonly influenced by noise from a mixture of sources, including optical sensing limitations and electronic functional failures, making image denoising an essential topic in low-level computer vision.
Failing to effectively remove noise can significantly reduce the performance of subsequent high-level tasks or scenes. 
For instance, this may affect content extracted from a low-light, noisy document capture or identification of pedestrians in adverse weather conditions for vision-based autonomous driving systems.
Furthermore, recent advancements in computational resources and specialized hardware such as image signal processors (ISPs), combined with the inherent limitations of optical systems, create a timely opportunity to advance denoising methods further.

Various foundational architectures have emerged, combining classical image processing principles with recent deep learning advances, which include convolutional neural networks (CNNs) \cite{zhang2017beyond}, Transformers \cite{chen2021pre}, Mambas \cite{zhu2024vision}, and Receptance Weighted Key Values (RWKVs) \cite{peng2023rwkv}. 
CNNs, often considered as an evolution of classical methods utilizing hand-crafted priors \cite{zheng2019mixed,laghrib2024image,he2010single}, remain their core idea of local window filtering. 
Despite advanced techniques such as non-local operations \cite{wang2018non}, large kernels \cite{ding2022scaling}, and architectures like U-Net \cite{ronneberger2015u}, CNNs struggle with oversmoothing in complex scenes and fail to handle distant information.
Transformers and Mambas introduce global modeling through self-attention (SA) mechanisms \cite{dosovitskiy2020image} and state space models \cite{zhu2024vision,liu2024vmamba}, respectively.
While these approaches achieve the state-of-the-art (SOTA) performance, they encounter challenges such as quadratic computational complexity in Transformers and causal-style dependencies in Mambas.
Alternative approaches \cite{jin2024linformer} tried to tackle these problems, but still lead to inconsistent modeling in local regions.

Reviewing prior work reveals two main issues: feature-level design and effective model architecture development.
Consider a scenario where a single convolution layer efficiently reduces noise in detailed regions but struggles in flat areas, leaving residual noise.
In such cases, the high-frequency components, often dominated by noise in flat regions, require attenuation rather than selective enhancement.
In addition, image information is spatially distributed across all directions. 
However, causal-style token mixing, where information is derived only from past tokens, introduces asymmetry in both weighting and positioning.
While this lack of symmetry may have minimal impact in natural language processing (NLP), it poses a significant challenge in image feature representation, where spatial symmetry is essential for accurate reconstruction and effective denoising.

In this study, we propose Context-guided Receptance Weighted Key-Value (\M), a novel model that addresses challenges in noise modeling and computational efficiency mentioned above. Our model provides superior real-world image denoising performance within limited resources, and the main contributions are as follows:
\begin{itemize}
    \item We propose \M, a novel model for image denoising that integrates a multi-view learning approach. Our design enhances the RWKV model by introducing the BiWKV mechanism, enabling full pixel-sequence computation superior to causal-style selection with linear complexity relative to sequence length.
    \item We introduce CTS mechanism to effectively model local noise correlations in image. Besides, we propose FMix module to selectively process frequency-domain information and attenuate noise. They together significantly improve denoising performance.
    \item Comprehensive experiments on real-world image denoising datasets demonstrate that \M~consistently outperforms SOTA methods. Its strong generalization ability is validated through testing on multiple unseen datasets, showcasing its efficiency and effectiveness for practical denoising applications.
\end{itemize}

\section{Related Works}
\subsection{Real-world Image Denoising}
Image denoising is a fundamental problem in image processing, with applications ranging from photography to medical imaging. 
Classical methods, such as BM3D \cite{dabov2007image}, rely on hand-crafted priors and assumptions about noise characteristics, achieving good results in controlled scenarios but struggle with complex noise patterns.
With advancements in deep learning, modern approaches like DnCNN \cite{zhang2017beyond} have emerged, focusing primarily on removing Additive White Gaussian Noise (AWGN). 
While effective for synthetic noise, these methods face significant challenges when applied to real-world noise, which is far more complex than AWGN due to spatial correlation, intensity variation, and limited paired data.

To address these issues, various methods have been proposed. 
Some rethink the imaging and noise generation processes, as seen in CBDNet and SCUNet \cite{guo2019toward,zhang2023practical}, while others adopt self-supervised pixel reconstruction techniques using blind-spot networks \cite{krull2019noise2void,lee2022ap}. 
Additionally, Gaussian denoisers have been adapted for real-world denoising by performing noise pattern corruption with shuffling techniques in advance \cite{zhou2020awgn,xiao2023random}.
Recent works have also explored novel image restoration architectures.
For instance, MambaIR \cite{guo2024mambair} leverages state space modeling, while Restormer \cite{zamir2022restormer} incorporates windowed self-attention, both demonstrating notable improvements in real-world image denoising.
Despite these advancements, limited attention has been given to integrating noise-specific priors into modern architectures while achieving a balance between computational efficiency and restoration quality.

\subsection{RWKV Models}
RWKV \cite{peng2023rwkv} was proposed as an efficient alternative to Transformers for NLP, particularly in Large Language Models (LLMs). 
RWKV introduces two main innovations: token shifting and WKV computation.
By incorporating token shifts in the preceding direction and concatenating shifted and non-shifted channels, the model can separate two tasks—predicting the next token, and accumulating and passing information from previous tokens.
The WKV computation improves upon the Attention-Free Transformer \cite{zhai2021attention} by employing trainable distance factors and enhanced integration of the current token to model token weights more precisely.
Additionally, it adopts an equivalent recurrent form for efficient inference. 
These features position RWKV as a strong competitor to CNNs and Transformers.

Recent studies, such as Vision-RWKV \cite{duan2024vision}, have showed the potential of RWKV-based models as vision backbones \cite{fei2024diffusion,zhou2024bsbp}. 
Two major improvements have been developed to adapt RWKV for vision tasks: quad-shift, a token-shifting strategy on 2D planes, and Bi-WKV, an attention mechanism with absolute positional bias to align with the symmetric nature of images.
While promising efforts like Restore-RWKV \cite{yang2024restore} have emerged, targeting all-in-one medical image restoration with Re-WKV and omni-shift, there remains limited exploration of how RWKV-based models can enhance performance in low-level vision tasks, such as real-world image denoising.

\section{Methodology}
\begin{figure*}[!t]
\centering
\includegraphics[width=\linewidth]{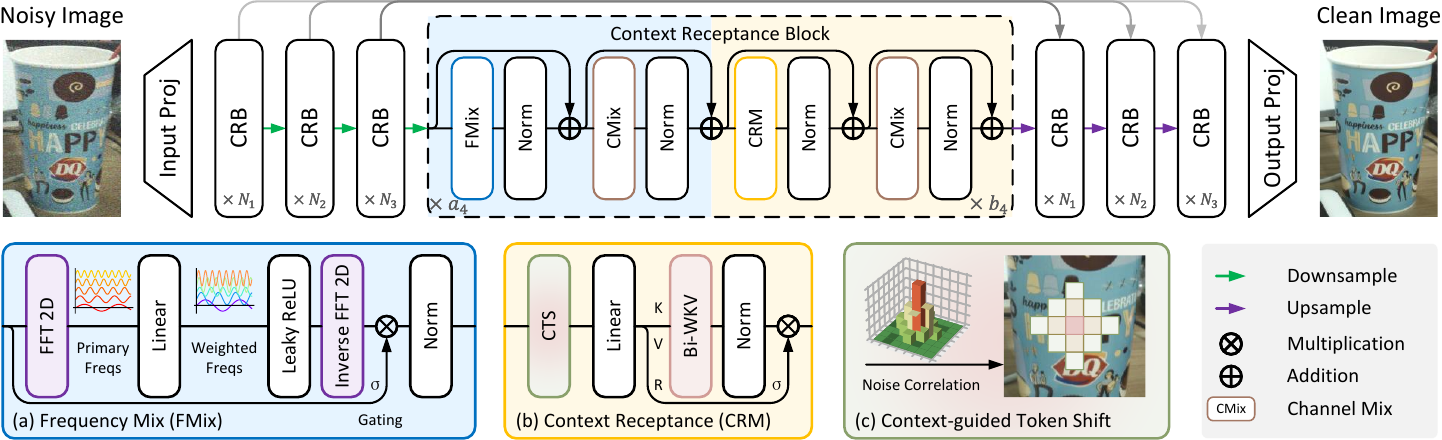}
\caption{Architecture of the proposed \M~model, and (a) Frequency Mix Module (FMix), (b) Context Receptance Module (CRM) and (c) Context-guided Token Shift (CTS).}
\label{fig:rwkvcr_arch}
\end{figure*}

In this section, the details of our proposed model \M~will be introduced. 
We will start with the model overview and gradually analyze the structure of CRB and FMix designed specifically for real-world noise modeling.

\subsection{Overall Architecture}
As illustrated in Figure~\ref{fig:rwkvcr_arch}, the proposed \M~model adopts a U-shaped hierarchical structure with long-skip connections to effectively capture both local and global features.
For a noisy input image $x \in \mathbb{R}^{H \times W \times 3}$, a $3 \times 3$ convolution is first applied to extract low-level features. The resulting feature map is then processed through four distinct stages of encoders and decoders following the U-shaped structure. After this process, the refined features are passed through a final projection layer to reconstruct the denoised output image.

Each encoder and decoder stage contains $N_k$ Context Receptance Blocks (CRBs). Each CRB consists of two types of modules for multi-view learning: $a_k$ Frequency Mix (FMix) Modules and $b_k$ Context Receptance Modules (CRMs), such that $a_k + b_k = N_k$ for $k = 1, 2, 3, 4$. The computation in a single CRB can be expressed as:
\begin{equation}
\begin{aligned}
    z_1&=\textrm{Norm}(\textrm{FMix}(x))+\alpha_1 x\,, \\
    z&=\textrm{Norm}(\textrm{CMix}(z_1))+\alpha_2 z_1,
\end{aligned}
\end{equation}
and 
\begin{equation}
\begin{aligned}
    y_1&=\textrm{Norm}(\textrm{CRM}(z))+\beta_1 z\,, \\
    y&=\textrm{Norm}(\textrm{CMix}(y_1))+\beta_2 y_1\,,
\end{aligned}
\end{equation}
where $x$ and $y$ denote the input and output features, respectively.
The Channel Mix (CMix) module is computed as follows:
\begin{equation}
\begin{aligned}
        r_c,k_c&=\textrm{CTS}(z)\,, \\
      \textrm{CMix}(z)& =\sigma(L(r_c))\odot\textrm{Norm}(\max(0,k_c^2))\,,
\end{aligned}
\end{equation}
where $\text{CTS}(\cdot)$ represents the Context-guided Token Shift mechanism, further detailed in Section~\ref{para:cts}.

\subsection{Context Receptance Module}

\begin{figure}[!t]
\centering
\includegraphics[width=\linewidth]{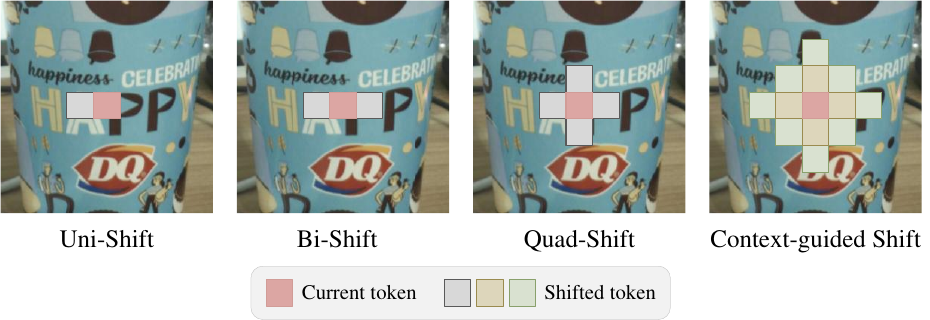}
\caption{Illustration of different token shift mechanisms.}
\label{fig:shift_visual}
\end{figure}

\begin{algorithm}[tb]
\caption{Context-guided Token Shift}
\label{alg:cts}
\textbf{Input}: input $x$, offset dictionary $D$, learnable weight $\omega$\\
\textbf{Parameter}: channel $C$\\
\textbf{Output}: shifted output $\textrm{CTS}(x)$
\begin{algorithmic}[1]
    \STATE Let $p_{\mathrm{sum}}=0,c=0,o=\textrm{zeros}(x.\textrm{shape})$
    \FOR {offset $p$ in $D$}
    \STATE calculate $p$'s Manhattan distance $d_p=d_m(p,0)$
    \STATE calculate offset $p$'s weight $w_p=1/d_p$
    \STATE $p_{\mathrm{sum}}\mathrel{{+}{=}}w_p$
    \ENDFOR
    \STATE calculate channel expansion factor $k =C/p_{\mathrm{sum}}$
    \FOR {offset $p$ in $D$}
    \STATE fill $o$ w/ shifted $x$: $o[c:c+k\cdot w_p]=x_p[c:c+k\cdot w_p]$
    \STATE $c=c+k\cdot w_p$
    \ENDFOR
    \RETURN $\textrm{CTS}(x)=\omega\cdot o+(1-\omega)\cdot x$
\end{algorithmic}
\end{algorithm}

\begin{table*}[!t]
  \centering
  \setlength{\tabcolsep}{4pt} 
  \renewcommand{\arraystretch}{0.9}
  \scalebox{0.9}{
  \begin{tabular}{l|*{5}{r}}
    \toprule
    Characteristics & Vanilla Attention & Window Attention & Linear Attention & State Space Model & BiWKV \\
    \midrule
    Operator Type & SA & SA & Linearized SA & Selective scan & Attention-free \\
    Inductive Bias & - & Locality & Low-rank (LR) approx. & Causality & Spatial symmetry \\
    Token Mixer & Dense & Blockwise sense & Lower-triangle (LT) & LT, blockwise LR & Almost dense \\
    Token Type & Patch (typ.) & Patch (typ.) & Patch (typ.) & Pixel (typ.) & Pixel (typ.) \\
    Local Interaction & - & Shifting window & - & Scan strategy & Token shifting \\
    Global Interaction & Direct & Arch-specific & Kernel-dependent & Hidden state & Recurrent state \\
    Complexity & Quadratic & Quadratic to win\_size & Linear & Linear & Linear \\
    \bottomrule
  \end{tabular}}
  \caption{Comparison of different operators.}
  \label{tab:comparison_op}
\end{table*}

\paragraph{Context-guided Token Shift.} 
\label{para:cts}
The shifting operations in the mixing processes enable selective accumulation of information from previous tokens to the current token.
Unlike NLP problems, pixels in an image often exhibit correlations with their neighbors in a centrosymmetric manner.
Real-world noise, in particular, correlate with a specific neighborhood structure.
Therefore, it is essential to identify this neighborhood shape to ensure that all significant pixels are fully considered while avoiding shifts over excessively large areas, which may introduce extra complexity in the implementation and reduce efficiency.
 
To address this issue, we propose a CTS mechanism, which assigns predefined weights to the most correlated pixels, as illustrated in Figure~\ref{fig:rwkvcr_arch}(c)\added{, and a visual comparison with existing token shift methods is provided in Figure~\ref{fig:shift_visual}}.
The colored masks in the figure depict individual pixels rather than image patches. 
Starting with the red central pixel, a context-guided region is defined as the equivalent reception field.
This region aligns with the specific neighborhood identified in \cite{wang2023lg} through Pearson's correlation analysis, which highlights the most correlated pixels with the central noise. 
The detailed algorithm of CTS is presented in Algorithm~\ref{alg:cts}.

\paragraph{Bidirectional WKV operation.} 
Unlike the self-attention mechanism, BiWKV operation employs a token-shift operation to achieve a weighted fusion of the feature map with its context-guided shifted version, generated using the CTS operation.
The fused feature map, denoted as $\textrm{CTS}(x)$, is then used to produce $r_1$, $k_1$, and $v_1$ through three linear projection layers:
\begin{equation}
\begin{aligned}
    r_1=\textrm{CTS}(x)W_r\,,k_1=\textrm{CTS}(x)W_k\,,v_1=\textrm{CTS}(x)W_v,
\end{aligned}
\end{equation}
where $W_r$, $W_k$, and $W_v$ are the weight matrices of the respective linear projection layers. 
Among these, $k_1$ and $v_1$ are utilized for the BiWKV computation, while $r_1$ serves as a gating mechanism after passing through a sigmoid activation function.
The output of a single CRM block is computed as:
\begin{equation}
\begin{aligned}
    \textrm{CRM}(x)=\sigma(r_1)\odot\textrm{Norm}(\textrm{BiWKV}(k_1,v_1))\,,
\end{aligned}
\end{equation}
where $\sigma$ represents the sigmoid function and $\odot$ is the Hadamard element-wise product. 
Denote the relative position bias between the $t$-th and $i$-th token as:
\begin{equation}
\begin{aligned}
    b_{t,i}=-(|t-i|-1)/ T\,,
\end{aligned}
\end{equation}
the computation of the BiWKV operation is given as follows:
\begin{equation}
\begin{aligned}
     &\textrm{BiWKV}(k_1,v_1)_t \\
    =&\frac
    {\sum_{i\ne t}\exp\left(b_{t,i}w+k_{1,i}\right) v_{1,i} 
    +\exp\left(u+k_{1,t}\right)v_{1,t}}
    {\sum_{i\ne t}\exp\left(b_{t,i}w+k_{1,i}\right) 
    +\exp\left(u+k_{1,t}\right)}\,,
\end{aligned}
\end{equation}
where $T$ represents the length of the processed sequence, $k_{1,i}$ and $v_{1,i}$ correspond to the $i$-th token of $k_1$ and $v_1$, respectively, and $u$ is a learnable parameter representing bonus of the current token.
The absolute valued design in $b_{t,i}$ ensures that tokens equidistant from the current token in both forward and backward directions are weighted equally, preserving the spatial symmetry of image planes.
Overall, the BiWKV operation computes a weighted sum of all tokens in $v_1$ along the token dimension. The weights are determined by a combination of symmetric relative position bias, the vector $k_1$, and the current token bonus controlled by the parameter $u$.

To justify the BiWKV operation for real-world image denoising, we compare it to several attention mechanisms and sequence models relevant to low-level vision.
These include the vanilla attention \cite{dosovitskiy2020image}, window attention \cite{liang2021swinir}, linear attention \cite{cai2023efficientvit}, state space model \cite{guo2024mambair}, and BiWKV.
Table~\ref{tab:comparison_op} summarizes a detailed comparison of these operations, focusing on key characteristics. Specifically, the `Token Mixer' row in the table reflects the shape of $L$ in the following equation, based on the structured masked attention framework \cite{dao2024transformers}:
\begin{equation}
\begin{aligned}
    y=f(L\circ (QK^\top))\cdot V\rlap{.}
\end{aligned}
\label{eq:struc_attn}
\end{equation}

\subsection{Frequency Mix Module}
Real-world images often contain spatially correlated noise, which complicates detail-rich areas. This noise spreads information across all frequencies, causing overlap in high-frequency components and resulting in either over-smoothing or incomplete noise removal.

To address this issue, we propose the FMix module, illustrated in Figure~\ref{fig:rwkvcr_arch}(a), designed to extract detailed frequency representations from high-level features using FFT.
Given a feature map $x\in\mathbb{R}^{h\times w\times c}$, a 2D FFT is applied:
\begin{equation}
\begin{aligned}
    x^F_{u,v,c}=\sum_{m=0}^{h-1} \sum_{n=0}^{w-1} x_{m,n,c}\exp\left({-2\pi i\left(\frac{ux}{h}+\frac{vy}{w}\right)}\right)\rlap{.}
\end{aligned}
\end{equation}
The extracted frequencies are linearly weighted, followed by activation with Leaky ReLU. The weighted frequencies are then passed through an inverse FFT (iFFT) to return to the spatial domain.
The resulting feature map is combined with the input via an element-wise product and normalized to produce the filtered feature map:
\begin{equation}
\begin{aligned}
    x^F&= \textrm{FFT}(x)\,,\\
    z&=\textrm{LReLU}(\textrm{Linear}(x^F))\,,\\
    \textrm{FMix}(x)&= \textrm{Norm}(\textrm{iFFT}(z)\cdot x)\rlap{.}
\end{aligned}
\end{equation}

\subsection{Loss Function}
The proposed \M~model is trained by minimizing the $L_1$ loss, which can be written as follows:
\begin{align}
    L_1(y,x^*)=\Vert y-x^*\Vert_1\,,
\end{align}
where $x^*$ is the ground truth and $y$ is the model output. The $L_1$ loss is capable of retaining fine details such as edges and textures with its linear penalty that other losses such as $L_2$ may smooth out, ensuring robustness to outliers and accurate reconstruction of the denoised image.

\section{Experiments}
\begin{table*}[!t]
\centering
\setlength{\tabcolsep}{6pt} 
\renewcommand{\arraystretch}{0.9} 
\scalebox{0.9}{
\begin{tabular}{l|r|rr|rr|rr|rr}
    \toprule
        \multirow{2}{*}{Methods} & \multirow{2}{*}{Params (M)} & \multicolumn{2}{c|}{SIDD} & \multicolumn{2}{c|}{ccnoise} & \multicolumn{2}{c|}{PolyU} & \multicolumn{2}{c}{Urban100GP} \\
        & & $\uparrow$PSNR & $\uparrow$SSIM & $\uparrow$PSNR & $\uparrow$SSIM & $\uparrow$PSNR & $\uparrow$SSIM & $\uparrow$PSNR & $\uparrow$SSIM \\ 
    \midrule
        BM3D \cite{dabov2007image} &
        - &
        29.97 & 0.679 &
        36.15 & 0.947 &
        37.40 & 0.957 &
        25.02 & 0.813 \\
        
        AP-BSN \cite{lee2022ap} &
        3.10 &
        36.74 & 0.889 &
        33.30 & 0.918 &
        36.46 & 0.947 &
        24.25 & 0.716 \\

        B2U \cite{wang2022blind2unblind} &
        1.96 &
        32.37 & 0.727 &
        35.72 & 0.938 &
        35.71 & 0.947 &
        27.39 & 0.847 \\
        
        DnCNN \cite{zhang2017beyond} &
        0.56 &
        26.21 & 0.604 &
        33.88 & 0.959 &
        36.11 & 0.960 &
        24.20 & 0.866 \\ 

        SwinIR \cite{liang2021swinir} &
        11.75 &
        33.70 & 0.864 &
        35.26 & 0.978 &
        37.14 & 0.977 &
        \underline{28.21} & 0.896 \\
    
        Uformer \cite{wang2022uformer} &
        50.88 &
        39.68 & 0.958 &
        36.02 & 0.979 &
        37.48 & 0.979 &
        26.88 & 0.885 \\

        ShuffleFormer \cite{xiao2023random} &
        50.53 &
        39.60 & 0.958 &
        35.88 & 0.978 &
        37.50 & 0.979 &
        27.89 & 0.900 \\

        Restormer \cite{zamir2022restormer} & 
        26.10 &
        \textbf{40.01} & 0.960 &
        \underline{36.33} & 0.981 &
        37.56 & 0.979 &
        28.18 & \underline{0.904} \\

        MambaIR \cite{guo2024mambair} & 
        26.78 &
        \underline{39.88} & \underline{0.960} &
        36.20 & \underline{0.981} &
        \underline{37.58} & \underline{0.980} &
        27.42 & 0.898 \\
        
        \M~(Ours) &
        20.19 &
        39.87 & \textbf{0.960} &
        \textbf{36.69} & \textbf{0.983} &
        \textbf{37.60} & \textbf{0.980} &
        \textbf{28.30} & \textbf{0.908} \\
    \bottomrule
\end{tabular}}
\caption{Quantitative results of real-world image denoising on SIDD, ccnoise, PolyU and Urban100GP.}
\label{tab:realdn_results_all}
\end{table*}

\begin{figure*}[htbp]
\centering
\includegraphics[width=\linewidth]{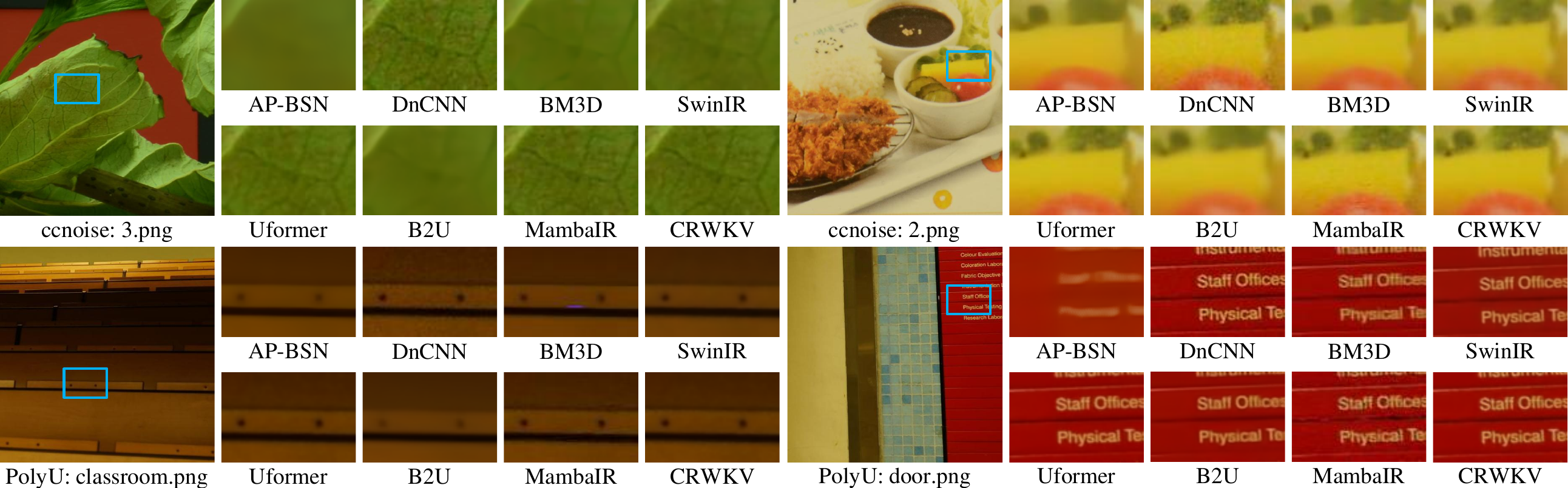}
\caption{Selected visual results on ccnoise and PolyU dataset.}
\label{fig:ccpolyu_visual}
\end{figure*}

\begin{figure}[t]
\centering
\includegraphics[width=\linewidth]{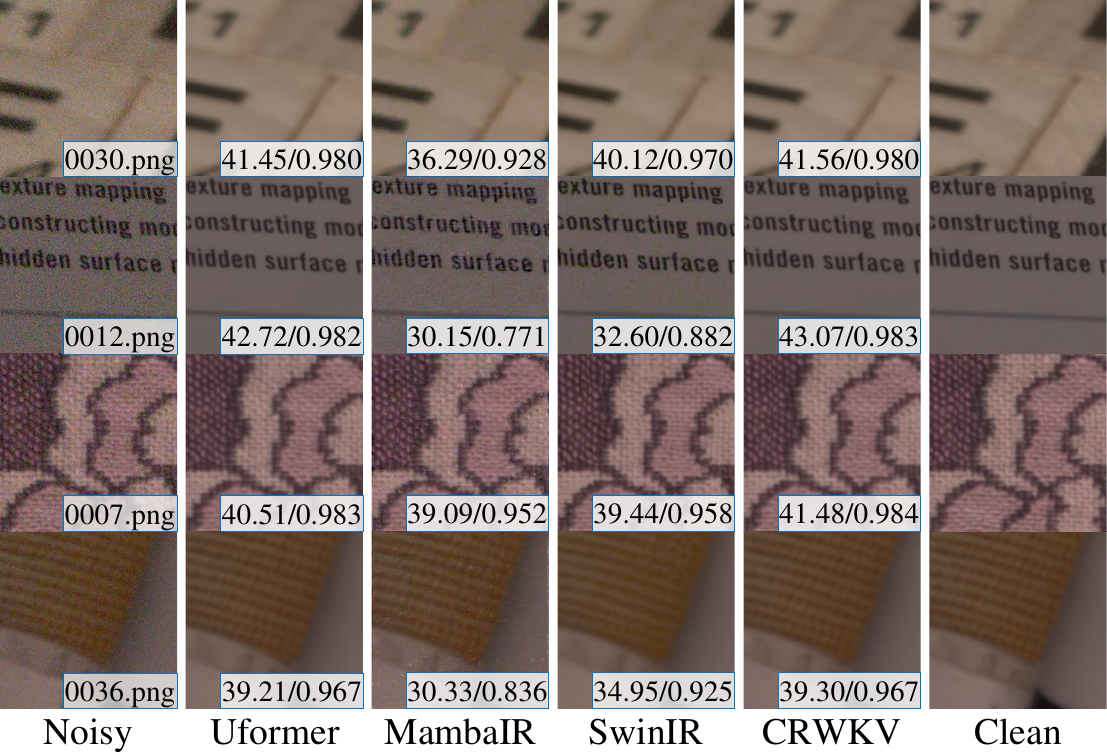}
\caption{Selected visual results on SIDD dataset.}
\label{fig:sidd_visual}
\end{figure}

\begin{figure}[t]
\centering
\includegraphics[width=\linewidth]{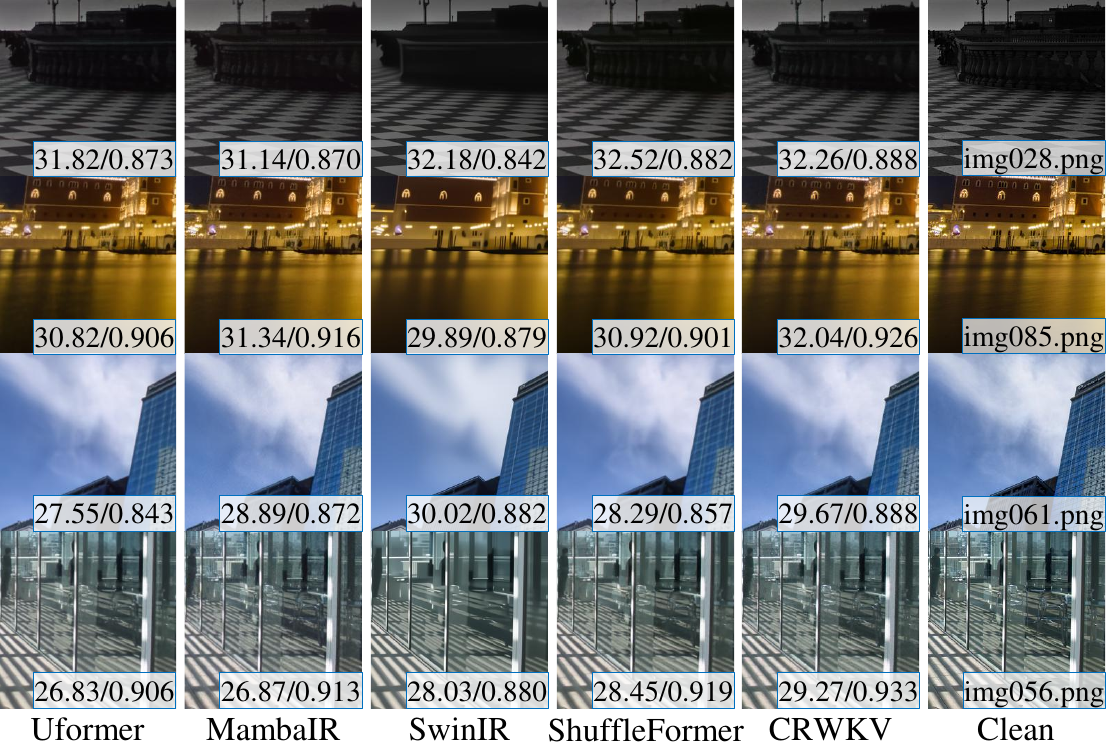}
\caption{Selected visual results on Urban100GP dataset.}
\label{fig:urban_visual}
\end{figure}

\subsection{Experiment Setup}
\paragraph{Datasets and metrics.}
We use the SIDD dataset \cite{abdelhamed2018high}, consisting of 320 real-world images, as the primary training set. 
From each high-resolution image ($256 \times 256$), we crop 300 non-overlapping slices, generating a total of 96,000 training samples. 
For testing, we evaluate on the SIDD, ccnoise \cite{nam2016holistic}, and PolyU \cite{xu2018real} datasets, all containing realistic noise. 
Additionally, we create a synthetic dataset, Urban100GP, by introducing mixed Additive White Gaussian Noise (AWGN) with $\sigma=10$ and Poisson noise to Urban100 \cite{huang2015single} to simulate real-world noise.
To quantify model performance, we use Peak Signal-to-Noise Ratio (PSNR) and Structural Similarity Index Measure (SSIM) as evaluation metrics.

\paragraph{Implementation details.}
During training, images are cropped to $128 \times 128$, and data augmentation techniques such as rotations ($90^\circ$, $180^\circ$, $270^\circ$) and random flipping are applied to enhance model robustness.
The training process is carried out with a batch size of $4$ for a total of $288,000$ iterations.
We use the AdamW optimizer with $\beta_1=0.9$ and $\beta_2=0.999$.
The learning rate starts at $3\times10^{-4}$ and is gradually reduced to $1\times10^{-6}$ after the $192,000$-th iteration. 
For model-specific configurations, the output channel size of the input projection is set to $48$. 
The depths of the four stages are empirically chosen as $L_1=3$, $L_2=L_3=4$, and $L_4=6$.
All experiments are conducted on a single NVIDIA RTX 4090 GPU, running Ubuntu 22.04 with PyTorch 2.5 as the software environment.

\begin{figure}[t]
\centering
\includegraphics[width=\linewidth]{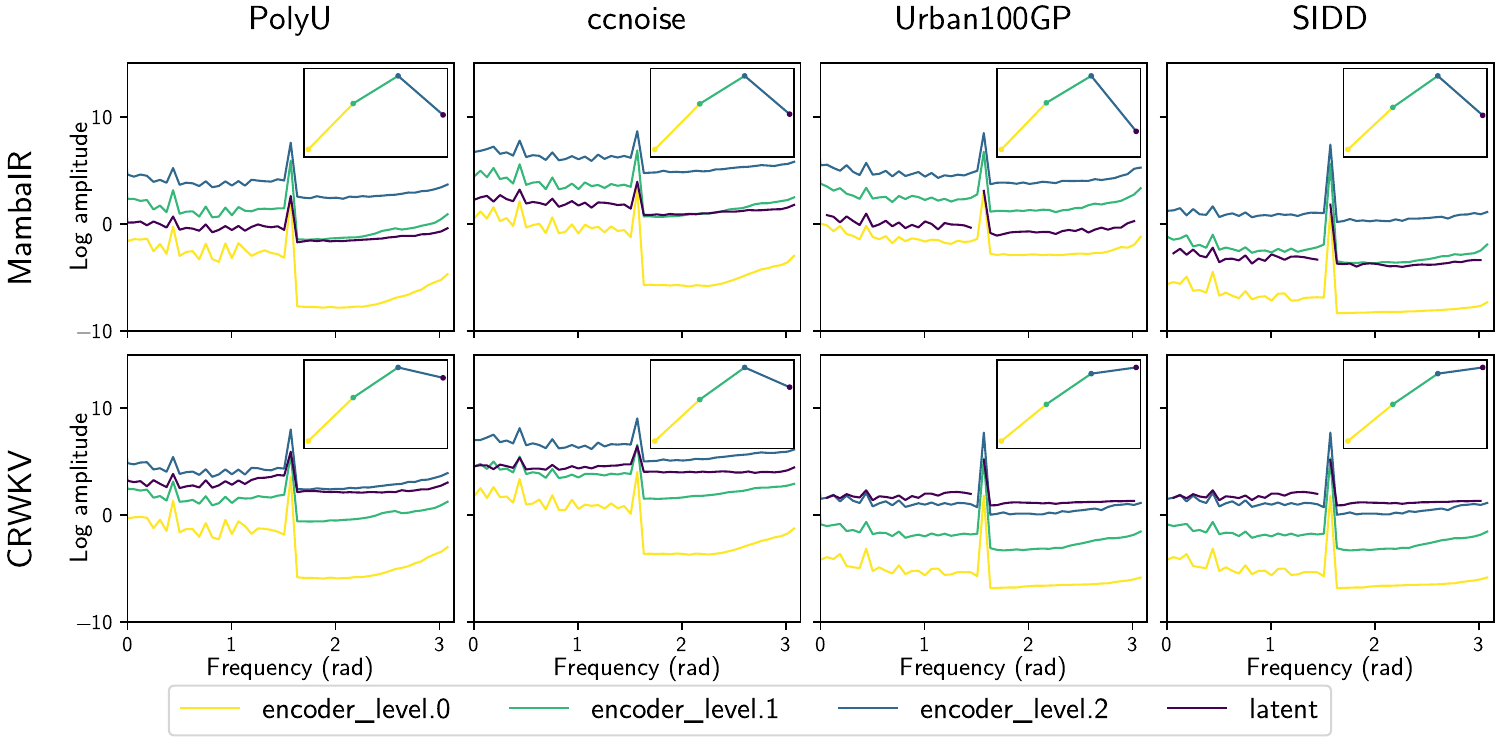}
\caption{Power spectrum of feature maps at different depths of MambaIR and \M~across various datasets. Lines with deeper colors represent deeper layers.}
\label{fig:layer_fig}
\end{figure}

\subsection{Comparison on Real-world Image Denoising}
The proposed FRWKV method was evaluated on real-world image denoising tasks against several SOTA methods, including BM3D \cite{dabov2007image}, AP-BSN \cite{lee2022ap}, B2U \cite{wang2022blind2unblind}, DnCNN \cite{zhang2017beyond}, Uformer \cite{wang2022uformer}, SwinIR \mbox{\cite{liang2021swinir}}, ShuffleFormer \cite{xiao2023random}, Restormer \cite{zamir2022restormer}, and MambaIR \cite{guo2024mambair}. 
The compared methods cover backbones including CNNs, Transformers and Mambas, functioning paradigms including supervised, self-supervised, and non-learning based. 

\paragraph{Quantitative comparison.}
Quantitative results on four datasets—SIDD, ccnoise, PolyU, and Urban100GP—are summarized in Table~\ref{tab:realdn_results_all}, along with the parameter count for each model.
Our model achieves superior performance across nearly all datasets and metrics.
The only exception is the PSNR metric on the SIDD dataset, where Restormer slightly surpasses our model by a margin of 0.1 dB. 
However, this minor advantage comes at the cost of a significantly larger model size, with Restormer requiring 30\% more parameters than \M.
On the ccnoise and PolyU datasets, the differences between the models' metrics are less significant compared to the SIDD dataset.
This can be attributed to the complex noise pattern in SIDD images, including noise introduced from multiple stages of the image processing pipeline.

Figure~\ref{fig:layer_fig} compares the power spectrum of feature maps between MambaIR and \M. 
A significant amplitude drop is observed in MambaIR between encoder\_layer.2 and latent, which is essential in semantic-level feature reconstruction. 
In contrast, \M~exhibits a much smaller amplitude reduction on PolyU and ccnoise while maintaining superior performance on Urban100GP and SIDD. 
This indicates that \M~is able to retain amplitude stably across layers, preserving and leveraging high-frequency information throughout the denoising process effectively.

\paragraph{Qualitative comparison.} 
On ccnoise and PolyU datasets (Figure~\ref{fig:ccpolyu_visual}), our model excels in preserving both fine details (leaf veins in ccnoise 3.png) and flat regions (ccnoise 2.png), and maintaining high-quality text fidelity (PolyU door.png). 
Other methods, such as B2U, struggle with edge preservation (PolyU classroom.png), while models like MambaIR leave visible noise residuals in flat regions (ccnoise 2.png).

For SIDD dataset (Figure~\ref{fig:sidd_visual}), visual comparisons showcase \M's superior ability to restore text and textures.
In the first two examples, our model successfully reconstructs text at varying scales, while competing methods struggle.
In the latter two examples, \M~recovers intricate textures such as those in the 0036.png, where other models like SwinIR fail to reproduce fine details.
The PSNR and SSIM values displayed in the lower-right corners further confirm the model's effectiveness.
For the Urban100GP dataset (Figure~\ref{fig:urban_visual}), \M~produces denoised images that strike a superior balance between preserving structural details and reducing noise artifacts.
For instance, the grid pattern and fine features in img056.png can be jointly restored with \M, outperforming other models.
Additionally, \M~produces the most realistic reconstruction of the reflective water surface, showcasing its ability to handle challenging scenarios.

Compared to other methods, \M~demonstrates significant improvements in restoring both flat regions and fine details, achieving smoother textures and sharper edges. 
Furthermore, evaluations on unseen datasets, such as ccnoise, PolyU, and Urban100GP, suggest strong generalizability of \M, as they were excluded from the training process.

\begin{table}[!t]
\centering
\setlength{\tabcolsep}{3pt} 
\renewcommand{\arraystretch}{0.95} 
\scalebox{0.80}{
\begin{tabular}{l|r|r|r|rr}
\toprule
    \multirow{2}{*}{Methods} & Params & FLOPs & Time & \multicolumn{2}{c}{SIDD} \\
    & (M) & (G) & (ms) & PSNR & SSIM \\ 
\midrule
    SwinIR \cite{liang2021swinir} & 11.75 & 253.46 & 170.33 & 33.70 & 0.864  \\ 
    ShuffleFormer \cite{xiao2023random} & 50.53 & 120.67 & 98.96 & 39.60 & 0.958 \\ 
    Uformer \cite{wang2022uformer} & 50.88 & 41.44 & 62.50 & 39.68 & 0.958 \\ 
    Restormer \cite{zamir2022restormer} & 26.10 & 35.24 & 47.40 & 40.01 & 0.960 \\ 
    MambaIR \cite{guo2024mambair} & 26.78 & 34.39 & 79.61 & 39.88 & 0.960 \\ 
    \M~(Ours) & 20.19 & 28.78 & 62.74 & 39.87 & 0.960 \\ 
\bottomrule
\end{tabular}}
\caption{Efficiency comparison with the SOTA methods.}
\label{tab:eff_comp}
\end{table}

\paragraph{Computational complexity.}
The model efficiency comparison results are summarized in Table~\ref{tab:eff_comp}.
On the ccnoise dataset, \M~outperforms Restormer by over 0.3 dB PSNR and achieves competitive performance with MambaIR, while utilizing only 83\% of the FLOPs.
Figure~\ref{fig:fig1}(b) illustrates GPU memory usage during inference for varying input image sizes, comparing \M~with state-space and full-attention architectures.
Notably, the Transformer-based Shuffleformer encounters memory limitations when the input size reaches 1024, whereas \M~requires only 40\% of the GPU memory.
Even with a comparable parameter count, \M~uses approximately 60\% of the memory consumed by state-space-based MambaIR.
This demonstrates that \M~strikes an optimal balance between denoising performance and computational efficiency, offering a practical solution for high-resolution and resource-constrained applications.

\subsection{Ablation Study}
\paragraph{Effectiveness of CRB module designs.}
To evaluate the effectiveness of the FMix module and CTS mechanism, we perform experiments on different configurations within the CRB.
In the first configuration, we remove FMix and replace it with a basic spatial-mix while omitting the CTS in the CRM and CMix modules.
In the second configuration, FMix is retained, but CTS is excluded. 
Finally, we evaluate the model's performance with partial (CRM-only or CMix-only) and full insertion of CTS.
The results suggest that removing FMix leads to a substantial performance degradation, with a PSNR drop of at least 0.40 dB on the SIDD dataset compared to the second configuration.
Similarly, omitting CTS from CRM and CMix results in suboptimal performance.
Introducing CTS yields consistent improvements, increasing PSNR by at least 0.15 dB on the ccnoise dataset and further boosting results on the SIDD dataset.
These findings validate the importance of FMix and CTS in enhancing denoising performance.

\paragraph{Effectiveness of shifting mechanisms.} 
As a guidance to the model's vast search space, shifting mechanisms include Uni-shift, Bi-shift, and Quad-shift are adopted previously.
To further evaluate the effectiveness of shifting window in CTS, we implement CTS(+), an extended version including pixels at a Manhattan distance of 3 from the central pixel additionally, covering a total of 16 neighboring pixels.
Table~\ref{tab:abl_shift} shows CTS outperforms CTS(+), suggesting the additional context introduces complexity without significant benefits.
Although learning a dynamic offset dictionary with deformable convolution is feasible, we choose a fixed offset dictionary to avoid overfitting to noise patterns in the training set and to reduce computational costs.
While the results show that a fixed offset dictionary achieves comparable performance, the need for learnable offsets for optimal results remains an open question.

\paragraph{Effectiveness of parameter settings.} 
To evaluate the roles of FMix and CRM in the model, we analyze the impact of parameters $a_k\,,b_k$ (where each CRB contains $a$ FMix modules and $b$ CRM modules), as shown in Table~\ref{tab:abl_ab}.
The findings can be summarized as follows: (1) A small number of FMix modules is optimal for low-level feature extraction and excessive usage disrupts early-stage modeling.
(2) Applying frequency domain analysis at the middle layers negatively impacts performance. This may be attributed to the limited ability of FMix to model mid-level features, which are less semantically structured.
(3) Incorporating frequency selection at deeper network layers significantly enhances performance. Deeper layers deal with semantically rich features that benefit more from frequency-domain processing.
Based on these findings, the optimal configuration for the CRB is $b_1=3\,,b_2=b_3=4$ and $a_4=6$.

\paragraph{Effectiveness of loss functions.}
Table~\ref{tab:abl_loss} presents a comparison of different loss functions on SIDD and ccnoise datasets. $L_1$ loss achieves the best performance, offering a strong balance between pixel-level fidelity and perceptual quality. In contrast, MSE loss, while converging faster, suffers from over-smoothing, leading to unsatisfied results. Charbonnier loss achieves comparable but slightly weaker results, likely due to its more complex convergence dynamics. PSNR loss, however, shows a significant performance drop on the ccnoise dataset, suggesting its limitations in generalizability.

\begin{table}[!t]
\centering
\setlength{\tabcolsep}{5pt} 
\renewcommand{\arraystretch}{0.9} 
\scalebox{0.9}{
\begin{tabular}{ccc|rrrr}
\toprule
    \multirow{2}{*}{FMix} & \multicolumn{2}{c|}{CTS} & \multicolumn{2}{c}{SIDD} & \multicolumn{2}{c}{ccnoise} 
    \\
    & CRM & CMix & PSNR & SSIM & PSNR & SSIM \\ 
\midrule
    \ding{56} & & & 39.28 & 0.954 & 36.45 & 0.981 \\
    \ding{52} & & & 39.70 & 0.957 & 36.50 & 0.982 \\
    \ding{52} & \ding{52} & & 39.75 & 0.957 & 36.64 & 0.983 \\
    \ding{52} & & \ding{52} & 39.74 & 0.957 & 36.62 & 0.982 \\
    \ding{52} & \ding{52} & \ding{52} & 39.87 & 0.960 & 36.69 & 0.983 \\
\bottomrule
\end{tabular}}
\caption{The effectiveness of CRB module designs.}
\label{tab:abl_comp}
\end{table}

\begin{table}[!t]
\centering
\setlength{\tabcolsep}{5pt}
\renewcommand{\arraystretch}{0.9}
\scalebox{0.9}{
\begin{tabular}{l|rrrr}
    \toprule
        \multirow{2}{*}{Shifting} & \multicolumn{2}{c}{SIDD} & \multicolumn{2}{c}{ccnoise} \\
        & PSNR & SSIM & PSNR & SSIM \\
    \midrule
        Uni-Shift & 39.57 & 0.957 & 36.46 & 0.982 \\
        Bi-Shift & 39.58 & 0.958 & 36.54 & 0.982 \\
        Quad-Shift & 39.74 & 0.958 & 36.50 & 0.982 \\
        CTS(+) & 39.79 & 0.958 & 36.60 & 0.982 \\
        CTS & 39.87 & 0.960 & 36.69 & 0.983 \\
    \bottomrule
\end{tabular}}
\caption{The effectiveness of shifting mechanisms.}
\label{tab:abl_shift}
\end{table}

\begin{table}[!t]
\centering
\setlength{\tabcolsep}{5pt}
\renewcommand{\arraystretch}{0.9}
\scalebox{0.85}{
\begin{tabular}{ccc|rrrr}
    \toprule
        \multirow{2}{*}{$k$} & \multirow{2}{*}{$L_k$} &\multirow{2}{*}{$(a_k,b_k)$} & \multicolumn{2}{c}{SIDD} & \multicolumn{2}{c}{ccnoise} \\
        & & & PSNR & SSIM & PSNR & SSIM \\
    \midrule
        \multirow{4}{*}{$1$} & \multirow{4}{*}{$3$} & $(3,0)$ & 39.72 & 0.957 & 36.13 & 0.935 \\
        & & $(2,1)$ & 39.82 & 0.958 & 36.60 & 0.980 \\
        & & $(1,2)$ & 39.83 & 0.958 & 36.57 & 0.925 \\
        & & $(0,3)$ & 39.87 & 0.960 & 36.69 & 0.983 \\
    \midrule
        \multirow{4}{*}{$2,3$} & \multirow{4}{*}{$4$} & $(4,0)$ & 32.77 & 0.779 & 22.91 & 0.727 \\
        & & $(2,1)$ & 34.63 & 0.789 & 30.71 & 0.824 \\
        & & $(1,2)$ & 38.56 & 0.908 & 34.90 & 0.884 \\
        & & $(0,4)$ & 39.87 & 0.960 & 36.69 & 0.983 \\
    \midrule
        \multirow{3}{*}{$4$} & \multirow{3}{*}{$6$} & $(6,0)$ & 39.87 & 0.960 & 36.69 & 0.983 \\
        & & $(3,3)$ & 39.84 & 0.958 & 36.63 & 0.982 \\
        & & $(2,4)$ & 39.85 & 0.957 & 36.59 & 0.982 \\
        & & $(0,6)$ & 39.84 & 0.957 & 36.65 & 0.983 \\
    \bottomrule
\end{tabular}}
\caption{The effectiveness of parameter settings.}
\label{tab:abl_ab}
\end{table}

\begin{table}[!t]
\centering
\setlength{\tabcolsep}{5pt}
\renewcommand{\arraystretch}{0.9}
\scalebox{0.9}{
\begin{tabular}{l|rrrr}
    \toprule
        \multirow{2}{*}{Loss function} & \multicolumn{2}{c}{SIDD} & \multicolumn{2}{c}{ccnoise} \\
        & PSNR & SSIM & PSNR & SSIM \\
    \midrule
        Charbonnier Loss & 39.86 & 0.959 & 36.67 & 0.983 \\
        MSE Loss & 39.66 & 0.957 & 36.53 & 0.983 \\
        PSNR Loss & 39.82 & 0.958 & 36.31 & 0.982 \\
        $L_1$ Loss & 39.87 & 0.960 & 36.69 & 0.983 \\
    \bottomrule
\end{tabular}}
\caption{The effectiveness of loss functions.}
\label{tab:abl_loss}
\end{table}

\section{Conclusion}
This work introduces the CRWKV model, a novel approach designed to tackle the challenges of noise modeling and computational inefficiency in real-world image denoising tasks. Key contributions include the CTS mechanism, which effectively captures local spatial contexts affected by noise, and the FMix module, which integrates semantic-level frequency-domain information through a multi-view learning process. By incorporating the BiWKV mechanism into the RWKV backbone, CRWKV achieves efficient pixel-sequence computation with linear complexity, overcoming the limitations of causal-style computation. These advancements enable CRWKV to effectively differentiate noise from complex scenes and separate high-frequency noise from structural details. Extensive experimental results demonstrate CRWKV's superior performance, both quantitatively and qualitatively, highlighting its robustness and practicality for real-world image denoising applications.

\section*{Ethical Statement}
There are no ethical issues.

\section*{Acknowledgment}
This work is supported by the Natural Science Foundation of Shandong Province (Grant No. ZR2023QF030).

\bibliographystyle{named}
\bibliography{ijcai25}

\end{document}